\input ./style/arxiv-general.cfg
\documentclass[aoas,MSNbibl,nameyear,dvips]{arximspdf}
\makeatletter
   \@ifpackageloaded{graphicx}{}{\usepackage{graphicx}}
\makeatother
\doi{10.1214/15-AOAS837}% Updated by VTEXPTS2LaTeX.exe, 28.09.2015
\volume{9}
\issue{4}
\pubyear{2015}
\firstpage{1997}
\lastpage{2022}
\docsubty{FLA}

\makeatletter
\newcommand{\rrvert}{\vert}
\newcommand{\rrVert}{\Vert}
\newcommand{\llvert}{\vert}
\newcommand{\llVert}{\Vert}
\renewcommand{\mid}{|}
\newcommand{\eqref}[1]{(\ref{#1})}
\newcommand{\argmin}{\operatorname{argmin}}
\newcommand{\tbb}{\tilde{\mathbf{b}}}
\newcommand{\by}{\mathbf{y}}
\newcommand{\X}{\mathbf{X}}
\newcommand{\bBeta}{\mathbf{B}}
\makeatother

\begin{document}
\begin{frontmatter}

\title{Regularized brain reading with shrinkage and~smoothing}
\runtitle{Regularized brain reading}

\begin{aug}
% Corresponding author: Rebecca Steorts - beka@cmu.edu% Updated by
%VTEXPTS2LaTeX.exe, 28.09.2015 13:34
\author[A]{\fnms{Leila}~\snm{Wehbe}\thanksref{M1,T1}\ead[label=e1]{lwehbe@berkeley.edu}},
\author[B]{\fnms{Aaditya}~\snm{Ramdas}\thanksref{M1,T2}\ead[label=e2]{aramdas@cs.berkeley.edu}},
\author[C]{\fnms{Rebecca~C.}~\snm{Steorts}\thanksref{M22,T3}\corref{}\ead[label=e3]{beka@stat.duke.edu}}
\and
\author[D]{\fnms{Cosma~Rohilla}~\snm{Shalizi}\thanksref{M11,T4}\ead[label=e4]{cshalizi@cmu.edu}}
\runauthor{Wehbe, Ramdas, Steorts and Shalizi}
\affiliation{University of California, Berkeley\thanksmark{M1},
Duke University\thanksmark{M22}\\
and
Carnegie Mellon University\thanksmark{M11}}
%\dedicated{}
\address[A]{L. Wehbe\\
Hellen Wills Neuroscience Institute\\
University of California, Berkeley\\
Berkeley, California 94720\\
USA\\
\printead{e1}}
\address[B]{A. Ramdas\\
EECS Department\\
University of California, Berkeley\\
Berkeley, California 94720\\
USA\\
\printead{e2}}
\address[C]{R.~C. Steorts\\
Department of Statistical Science\\
Duke University\\
Durham, North Carolina 27708\\
USA\\
\printead{e3}}
\address[D]{C.~R Shalizi\\
Department of Statistics\\
Carnegie Mellon University\\
5000 Forbes Ave.\\
Pittsburgh, Pennsylvania 15213\hspace*{16pt}\\
USA\\
\printead{e4}}
\end{aug}
\thankstext{T1}{Supported by NIH Grant 5R01HD075328-02 and by DARPA
Grant FA8750-13-2-0005.}
\thankstext{T2}{Supported by NSF Grant IIS-1247658.}
\thankstext{T3}{Supported by NSF Grants SES1130706 and DMS-10-43903 as
well as the John Templeton Foundation.}
\thankstext{T4}{Supported by NIH Grant \#2 R01 NS047493 and by NSF
Grant DMS-12-07759.}

% HISTORY:
%
\received{\smonth{1} \syear{2014}}% Updated by VTEXPTS2LaTeX.exe,
%28.09.2015 13:34
%
\revised{\smonth{12} \syear{2014}}% Updated by VTEXPTS2LaTeX.exe,
%28.09.2015 13:34

% ABSTRACT
%
\begin{abstract}
Functional neuroimaging measures how the brain responds to complex
stimuli. However, sample sizes are modest,
noise is substantial, and stimuli are
high dimensional. Hence, direct estimates are inherently imprecise and call
for regularization. We compare a suite of approaches which regularize via
\textit{shrinkage}: ridge regression, the elastic net (a generalization of
ridge regression and the lasso), and a hierarchical Bayesian model
based on
small area estimation (SAE).
We contrast regularization with \textit{spatial smoothing}
and combinations of smoothing and shrinkage.
All methods are tested on
functional magnetic resonance imaging (fMRI) data from multiple
subjects participating
in two different experiments related to reading, for both predicting
neural response
to stimuli and decoding stimuli
from responses. Interestingly, when the regularization parameters are
chosen by
cross-validation independently for every voxel, low/high regularization
is chosen in
voxels where the classification accuracy is high/low, indicating that
the regularization
intensity is a good tool for identification of relevant voxels for the
cognitive task.
Surprisingly, {all} the regularization methods work about
equally well, suggesting that beating basic smoothing and shrinkage
will take
not only {clever} methods, but {also careful} modeling.
\end{abstract}

% KEYWORDS
\end{frontmatter}

\setcounter{footnote}{4}
%s1 #&#
\section{Introduction}

A major goal of functional brain imaging is to relate activity levels in
various parts of the brain to differences in stimuli. Typical fMRI experiments
measure activity in tens of thousands of volume elements called voxels
(i.e., ``volume pixels'') within the brain, over $10^{2}$--${10}^{3}$ time steps,
while realistic stimuli vary on hundreds or thousands of dimensions
(see Section
\ref{secscientific-and-statistical-background}). Moreover, neuroscientists
want to study heterogeneity across the brain in responses to stimuli,
discounting noisy variations. Taking each voxel on its own, estimates of
response functions are inherently imprecise due to the level of the
noise and
the high dimensionality of the problem. In statistics, such estimation
problems are addressed by regularization, especially shrinking estimates
toward a reference value (e.g., 0). While shrinkage statistically stabilizes
parameters estimates, this may or may not help achieve the scientific
aim of
better understanding of the organization of the brain. Due to this, we examine
whether, and how, common regularization techniques serve the
inferential goals
of cognitive neuroscience.

Because such an investigation cannot be done abstractly, we study the behavior
of four different methods of regularizing linear regression, in two experiments
related to different aspects of reading. Three of our methods
regularize by
shrinkage: ridge regression; the elastic net, which generalizes both
ridge and
the lasso; and a hierarchical Bayesian (HB) model, developed for small area
estimation (SAE). Ridge regression, the lasso, and the elastic net exemplify
modern high-dimensional frequentist statistics, based on penalized
optimization. The SAE model is an instance of the Bayesian approach
increasingly used in neuroscience [\citet
{genovese2000,lee2011,pillow2013}],
where a hierarchical process generates the parameters. Our fourth
method of
regularization smooths the data over spatial regions. We also consider
combinations of shrinkage and spatial smoothing, developing a novel
decision-theoretic method for smoothed SAE in the spirit of \citet
{louis1984}
and \citet{datta2011}. All methods were compared to the
performance of
unregularized ordinary least squares (OLS) regression.

As mentioned, we evaluated our methods on two experiments: one studying the
representation of the meaning of individual word-picture pairs (\textup{E1}, Section
\ref{sece1-explained}), and the second studying story comprehension
(\textup{E2}, Section
\ref{sece2-explained}). The two experiments differed in their subject pool,
in the nature of the stimulus (independent, randomly presented word
picture-pairs vs. consecutive words of a real story that requires the
maintenance of a complex context representation), in how long each
stimulus was
presented (10~s vs. 0.5~s), and in the nature
of the
appropriate analysis (static vs. time series). Findings about regularization
methods common to both experiments are unlikely to be artifacts of just one
experiment.

Surprisingly, despite their different rationales and inner workings,
\textit{all} of our methods of regularization gave very similar out-of-sample
performances in both experiments. All achieved low mean-squared-errors in
predicting neural responses to stimuli, and high accuracy in
classifying novel
stimuli based on neural response (Section~\ref{secaccuracy}). They improved
in both
respects over unregularized OLS, though only slightly. They produced very
similar parameter estimates, especially ridge regression and SAE, a connection
explained in Section~\ref{secSAE}. They showed a consistent pattern of how much
estimates in different parts of the brain were regularized---they imposed
more shrinkage or more smoothing in areas of low signal strength. This
suppression of ``noisy'' brain regions is perhaps their greatest
advantage over
OLS (Section~\ref{secpatterns}). This indicates that single-voxel
regularization could
be viewed as a detector for informative brain regions because it allows
predicted brain activity to be different from zero only in the informative,
less noisy regions. The near-equal performance of all regularizers
means that
choices between them must be based on considerations such as
computational cost
(Section~\ref{seccomputational}) and/or biological plausibility. Improving on
these outcomes must come from better biological modeling and not more clever
general-purpose statistical methods.

The rest of this paper proceeds as follows. We present the necessary
neuroscientific background in Section~\ref
{secscientific-and-statistical-background} and summarize the two data sets
used in this paper (Sections~\ref{sece1-explained}~and~\ref{sece2-explained}). We describe the details of our methods in
Section~\ref{secmethods}, provide in-depth results in Section~\ref
{secresults}, and
conclude with a discussion in Section~\ref{secdiscussion}.

%s1.1 #&#
\subsection{Neuroscientific background}
\label{secscientific-and-statistical-background}

Cognitive neuroscientists use functional magnetic resonance imaging to study
how the brain implements cognition. FMRI specifically measures ``hemodynamic
response,'' the change in blood oxygen levels after neural activity, as
a proxy
for information processing [\citet{ashby2011}]. Let $y_{vt}$ be
the measured
activity at voxel $v$ and discrete time-point $t$. While these
measurements are
often smoothed spatially to reduce noise, most analyses involve running a
separate regression of each voxel against stimulus features
[\citet{ashby2011}, Chapter~5].

In experiments with static stimuli that are presented with enough time in
between different instances to make them effectively independent, one could
make the design matrix of the regression simply contain stimulus
features, and
use an appropriate time window to obtain a single average response
$y_{vt}$ for
each stimulus~$t$. In such cases, for precision there are usually multiple
repetitions of each stimuli. On the other hand, in experiments with dynamic
stimuli, the contents of the design matrix for such regressions are
dictated by
the fact that the hemodynamic response has a long time latency, typically
peaking about six seconds after stimulus onset. The time-courses of stimulus
features are convolved with a kernel function modeling the hemodynamic
response, resulting in a time-varying set of covariates in the design matrix.
For each $v$, $y_{vt}$ is regressed against these covariates.

Statically or dynamically, it is typically assumed that voxels which are
found to have statistically significant regression coefficients are actually
involved in processing the stimuli. (We return to this assumption in Section
\ref{secspotlight}.) The focus on statistical significance, and the
fact that
there are usually more points than covariates, explains the popularity of
unregularized OLS [\citet{ashby2011}, Chapter~5]. Spatial
information is not
often utilized, but region information is sometimes used to threshold
significance maps, searching for contiguous blobs of significant voxels
[\citet{smith2004overview}].

More recently, multivariate pattern analysis has used information from multiple
voxels to decode underlying cognitive states. In discriminative models, fMRI
images are fed as input to a classifier, which attempts to ``reverse infer''
the stimulus or the state of the brain. Some of these methods, especially
discriminative Bayesian models, take advantage of the spatial
smoothness of the
fMRI image [\citet{norman2006,pereira2009,pillow2013}]. Our interest,
however, lies in generative models that can both predict the fMRI
images that
arise in response to new stimuli and decode stimuli from responses.
Discriminative models can only decode, whereas generative models aim
for a more
complete understanding of neural dynamics [\citet{naselaris2011encoding}].

We use fMRI data from two experiments, the first static and the second dynamic,
for both predicting neural response to stimuli as well as decoding
stimuli from
responses. While both involve reading, the two experiments probe reading
differently, which we briefly describe in Sections~\ref
{sece1-explained} and
\ref{sece2-explained}. In both experiments, the data is analyzed by using
feature representations of the stimuli and then expressing brain
activity as a
function of these stimulus features; this idea was introduced in
\citet{mitchell2008predicting}. The use of feature
representations allows the
experimenter to predict brain activity for a novel \emph{unseen}
stimulus, by
multiplying regression coefficents learned on old \emph{seen} stimuli
by the feature
representations of the new stimulus. Thus, the model that is learned
can be
assessed in terms of how well it generalizes its prediction to unseen stimuli.

%s1.2 #&#
\subsection{Experiment 1: Visual features of word-picture combinations}
\label{sece1-explained}

The first experiment (\textup{E1}) scanned native English speakers as they
looked at
word-picture combinations, specifically sixty concrete nouns (e.g., ``apple'',
``car''), accompanied by black-and-white line drawings of those objects\break
[\citet{mitchell2008predicting}]. All nine subjects were exposed
six times each
to all sixty word-picture stimuli, varying in order. Here the latency
of the
hemodynamic response was handled by averaging the activity acquired 4--8
seconds after stimulus onset, resulting in a single brain image per
subject per
stimulus per exposure. The six repetitions of each stimulus are themselves
averaged together (within subjects) in the data set.\footnote{Data were
obtained from
\surl{http://www.cs.cmu.edu/\textasciitilde fmri/science2008/data.html},
accessed in November 2013.}

Each voxel was 3.125~mm${}\times3.125$~mm${}\times 6$~mm,
and every subject's brain contained $\approx$21,000 voxels. The subjects'
brains were morphed into the same anatomical space, although exact
overlap is
not achieved due to anatomical differences. The voxels are divided into 90
``regions of interest'' (ROIs), generally believed to be anatomically and
functionally distinct [\citet{tzourio2002automated}]. The ROIs
vary greatly in
size, from about 20 to about 800 voxels. For ROIs covering a large
volume of
the brain, the spatial smoothness we hope to exploit is washed out. To counter
this, and achieve uniformity of size, we divided ROIs that had more
than 200
voxels in half along their largest dimension ($x$, $y$, or $z$
coordinate). This
was repeated as necessary until all regions had 200 voxels or less.
After this,
we had 191 ROIs.

We used eleven features related to the visual properties of the stimuli (e.g.,
``amount of white pixels on the screen'', ``2D aspect ratio''). These
annotations were provided to us by the authors of \citet
{sudre2012tracking}, who
used the same stimulus set for a different experiment. The original experiment
reported these features as ordinal variables on a five-point scale. We
selected these features since they represent a fairly coherent set of
precisely measured aspects of the stimuli, ones whose processing is
well understood neurobiologically [\citet{shepherd1994}]. For the
same reasons,
we did not use the many other features also measured in the experiment which
are related to semantic or physical properties of the stimuli (e.g.,
``Is it
manmade?'', ``Can I hold it in one hand?''), as manually rated on the same
five-point scale by workers on Amazon's Mechanical Turk crowdsourcing system
[\citet{sudre2012tracking}].

%\textcolor{red}{I am sorry for not noticing this before, but really, {
%\em all}
% the features were ordinal? Unless the variables in our model are
%really
% indicators for the different levels, a linear model is pretty
%extravagant
% here. We need to clarify this. CRS; I agree very much as well and
%apologize for not catching this too. RCS -
% LW: I think Aadi, Beka and I discussed this point by email and agreed
%it was ok.}

In summary, the data (\textup{E1}) consists of sixty words, represented by eleven
features each, and their associated average voxel activity across nine
subjects.

%s1.3 #&#
\subsection{Experiment 2: Textual features in narrative comprehension}
\label{sece2-explained}

Our second experiment deals with the response to dynamic textual
features in a
narrative comprehension task [\citet{wehbe2014}]. Eight subjects
read Chapter~9
of {\em Harry Potter and the Sorcerer's Stone} [\citet
{rowling2012harry}] while
in the fMRI scanner. In order to know exactly when each word was
processed by
the subjects, only one word from the text was shown at a time, on the
center of
a screen, each word being projected for $0.5$ seconds. The sampling
rate of
fMRI acquisition was $2$ seconds per observation, hence, four
consecutive words
were read during the time it took to scan the whole brain once. The experiment
lasted 2710 seconds in total, giving us 1355 full brain scans.\footnote
{Data is available at
\surl{http://www.cs.cmu.edu/\textasciitilde fmri/plosone}.}

Spatially, the voxels were $3\mbox{ mm}\times3\mbox{ mm}\times
3\mbox{ mm}$, somewhat smaller than in \textup{E1}, and every subject's brain contained
$\approx$29,000 voxels. As done in \textup{E1}, the brains were morphed into a common
space and divided into ROIs, and we further subdivided excessively
large ROIs.

As in \textup{E1}, we again look at only features related to the visual
properties of
the stimuli. Since the visual stimulus being received by the subject at any
time is just a word printed on a screen, standardized in color, font,
etc., we
focus on a single quantitative textual feature which is comparable across
words, namely, their length in letters. Each observation spans four
words and,
hence, we used both the mean and the standard deviation of the length
of the
presented words as our features. To account for the latency and
persistence of
the hemodynamic response, the stimulus features at time $t$ are used as
regressors for the activity at times $t+1$ through $t+4$. As before, we discard
many of the features from \citet{wehbe2014} relating to different
kinds of
semantic properties of the stimuli, like parts-of-speech tags (noun,
verb, etc.)
as well as other aspects of the story (characters, suspense, etc.) to maintain
consistency in the paper and comparability with \textup{E1}.

In summary, the data (\textup{E2}) consist of two time series: (1) the mean and standard
deviation of word lengths in every two-second interval and (2) the associated
time series of voxel activities across eight subjects.

\subsubsection*{Contrast between \textup{E1} and \textup{E2}}
\textup{E1} probes the processing of static
visual stimuli in a rather simple (even artificial) reading task. In contrast,
\textup{E2} deals with dynamic, textual features in a narrative comprehension task.
While the serial presentation of words is rare outside of the
laboratory, the
words were presented at a comfortable rate, and the subjects were previously
asked to practice reading in this serial fashion [\citet
{wehbe2014}], making the
overall setting much closer to ``ecological validity'' than is \textup{E1}. Common
findings about the properties and performance of statistical methods across
such different settings are very unlikely to be artifacts of a {particular}
experiment. We now turn to the description of the methods applied to
both \textup{E1}
and \textup{E2}.

%s2 #&#
\section{Methods}
\label{secmethods}

In previous analyses of both experiments
[\citet{mitchell2008predicting,wehbe2014}], the neural response
to reading a word
was modeled as a linear combination of the word's features. While such
linear models are ubiquitous in fMRI data analyses [\citet
{ashby2011}], they
have little biological basis. Nevertheless, any smooth model can be locally
approximated by a linear regression over a sufficiently small domain,
where the
range of the feature variables here is fairly small. Plotting actual responses
against linear fits shows that the latter are reasonable in these experiments
(Figure 4 of the Supplementary Article [\citet{suppA}]). Hence, we follow
the existing literature in
using linear models, and explore multiple ways of fitting and
regularizing them---OLS, ridge regression, the elastic net, and a hierarchical Bayesian model
from small area estimation (SAE). We then consider including the
effects of
combining these techniques with various forms of spatial smoothing.
Section~\ref{secevaluation} outlines our evaluation criteria for
models and their
regularizations, by their ability to both predict neural activity from stimuli
and to reconstruct stimuli from activity.

%s2.1 #&#
\subsection{Notation and model specification}

%\textcolor{red}{$\beta_v$ was being doubly used, which is very
%confusing and I do not think we ever use the stacked version in the
%rest of the paper. I have slightly changed the notation (everyone see
%if they agree with this change); RCS}

We introduce notation consistent throughout the paper and note that
we refer to real-valued variables by lowercase letters without boldface,
vectors as boldfaced lowercase letters, and matrices in boldfaced uppercase.

In the linear model for static experiment \textup{E1}, the average hemodynamic
response $y_{vt}$ of voxel $v$ (for $v=1,\ldots,V$ for $V\approx
21{,}000$, varies
per subject) to the stimulus, a word, and its associated image,
displayed at
time $t$ (for $t=1,\ldots,T$ for $T=60$), is a linear combination of stimuli
features denoted by the $P$-dimensional feature vector $\mathbf{x}_t$
(for $P=11$),
\[
y_{vt} = \mathbf{x}_t^\top{\bolds{
\beta}}_v + \varepsilon_{vt},
\]
where ${\bolds{\beta}}_v$ is the $P$-dimensional regression
coefficient vector of $v$
and $\varepsilon_{vt}$ is mean-zero noise for voxel $v$ at time $t$, with variance
$\sigma^2_v$, combining measurement error corrupting our observation with
fluctuations and the effects of specification error. Finally, we assume that
the $\varepsilon_{vt}$ has a Gaussian distribution. More succinctly, we will
stack the $\mathbf{x}_t$s into a $T \times P$ matrix $\X$, and for
each voxel $v$,
write its activity over the course of \textup{E1} as a $T$-dimensional vector
$\by_v$.

For dynamic experiment \textup{E2}, the activity $y_{vt}$ of voxel $v$ (with $V
\approx 29{,}000$, varies per subject) at time $t$ (with $T=1355$) is
modeled as
a linear function of the history of the stimulus, a continuous story, whose
visual features are represented here as a time-series of two-dimensional
vectors $\mathbf{x}_t$,
\[
y_{vt} = \sum_{k=1}^{h}{
\mathbf{x}^\top_{t-k} {{\bolds{\beta }}_{v,k}}} +
\varepsilon_{vt},
\]
where $h$ represents how long the hemodynamic response to a stimulus
persists and ${\bolds{\beta}}_{vk}$ captures how the hemodynamic
response at voxel $v$
depends on the $k$th previous set of four words. We note that
the mean and standard deviation of the word length of the $t$th set of four
words is presented during the $t$th brain scan.
We do not include $k=0$
because we assume the time window when a stimulus is presented is too
early to
see a significant response of the voxel. As noted in Section~\ref
{sece2-explained}, we set $h=4$ here, meaning that the voxel activity at
any time $t$ is only affected by the preceeding 8 seconds (16 displayed
words). We also require ${{\bolds{\beta}}_{v,0}} = 0$, meaning that
there is a lag of two
seconds before the hemodynamic response is seen. At any time $t$, the
latest set of four words (taking 2 seconds) captured by $x_{vt}$ does not
play a role in the latest activity $y_{vt}$.

This may be put in a form more similar to the static case by regressing
$\by_{v}$ on the vector obtained by concatenating $ \mathbf{x}_t,
\mathbf{x}_{t-1}, \mathbf{x}_{t-2},
\mathbf{x}_{t-3}$ into a single $P$-dimensional feature vector $\bar
\mathbf{x}_t$ (for $P=8$).
We can similarly concatenate the regression coefficients for this concatenated
feature vector to get a $P$-dimensional regression vector~$\bar{\bolds
{\beta}}_v$. We
overload notation to refer to $\bar\mathbf{x}_t$ and $\bar{\bolds
{\beta}}_v$ as $\mathbf{x}_t$ and
${\bolds{\beta}}_v$, since from this point the methods apply to both
static and dynamic
settings.

In both cases, the residual sum of squares is
\[
\mathrm{RSS}_v = \sum_{t=1}^T
\bigl(y_{vt} - \mathbf{x}_t^\top{\bolds{\beta
}}_v\bigr)^2 = \llVert \by_{v} - \X{\bolds{
\beta}}_v\rrVert _2^2, %
\]
where $\llVert  \cdot\rrVert  ^2_2$ is the squared Euclidean norm. OLS estimates ${\bolds
{\beta}}_v$
by minimizing the in-sample RSS, giving $\hat{{\bolds{\beta}}}_v =
(\X^\top\X)^{-1}
\X^\top\mathbf{y}_{v}$. The\vspace*{1pt} covariance of the estimates, in a fixed
design, is
$\sigma^2_v (\X^\top\X)^{-1}$.

%s2.2 #&#
\subsection{Ridge regression and elastic net}

We now review both ridge regression and elastic net, giving the Bayesian
counterparts to both. Ridge regression stabilizes OLS estimates via a penalty
term [\citet{hoerl1970}]. Specifically, the ridge estimator solves
%e1 #&#
%
\begin{equation}
\hat{{\bolds{\beta}}}^{R}_{v} = \mathop{\argmin}_{{\bolds{\beta}}_v}
\mathrm{RSS}_v + \lambda_v\llVert {\bolds{\beta}}_v
\rrVert _2^2. \label{eqnridgedefined}
\end{equation}
Equivalently, ${\bolds{\beta}}^R_v$ is constrained to be small, $\llVert
{\bolds{\beta}}^R_v\rrVert  _2^2 \leq
c$, for some $c > 0$. The tuning parameter $\lambda_v$ controls the
degree of
regularization. The ridge approach has been used before in neuroimaging with
the {\em same} $\lambda$ for all voxels [\citet{mitchell2008predicting}].
Importantly, in Section~\ref{secresults}, we show that tuning $\lambda
$ {\em separately for each voxel} improves classification and prediction and
provides valuable information about neural organization.

While ridge regression was developed from a frequentist perspective, it
has a
well-known Bayesian interpretation [\citet{hastie2001}]. By
imposing a Gaussian
prior on ${\bolds{\beta}}_v$ with prior precision $\lambda$, we find
%e2 #&#
%
\begin{eqnarray}
\label{eqnridgeasbayes} y_{vt}| \mathbf{x}_t,{\bolds{
\beta}}_v & \stackrel{\mathrm{ind}} {\sim}& N\bigl(\mathbf{x}^\top_t{
\bolds{\beta}}_v, \sigma^2_v\bigr),
\nonumber\\[-8pt]\\[-8pt]\nonumber
{\bolds{\beta}}_v & \stackrel{\mathrm{i.i.d.}} {\sim} & N(0,1/
\lambda_v \mathbf {I}).
\nonumber
\end{eqnarray}
Under the formulation in (\ref{eqnridgeasbayes}), the posterior mode
coincides exactly with the solution to \eqref{eqnridgedefined}. The solution
to both formulations has a closed form:
\[
\hat{{\bolds{\beta}}}^{R}_{v,\lambda_v} = \bigl(\X^\top\X+
\lambda _{v} \mathbf{I}\bigr)^{-1} \X^\top
\by_{v}.
\]
The covariance is $\sigma^2_v (\X^\top\X+ \lambda_v
\mathbf{I})^{-1} \X^\top\X(\X^\top\X+ \lambda_v \mathbf
{I})^{-1}$, in a
fixed-design regression.

The {\em elastic net} of \citet{zou2005} generalizes ridge
regression and the
lasso of \citet{tibshirani1996}:
\[
\hat{{\bolds{\beta}}}^{\mathrm{EN}}_{v} = \mathop{\argmin}_{{\bolds{\beta}}_v}
\mathrm{RSS}_v + \lambda_{1v} \llVert {\bolds{\beta}}_v
\rrVert _1 + \lambda_{2v}\llVert {\bolds{
\beta}}_v\rrVert _2^2. %
\]
Setting $\lambda_{1v} =0$ recovers ridge regression, and $\lambda_{2v}=0$
recovers the lasso. The $L_1$ penalty makes $\hat{{\bolds{\beta
}}}_v^{\mathrm{EN}}$ sparse,
shrinking coefficients on superfluous variables to zero, while the $L_2$
penalty alone favors small but nonzero coefficients. Again, previous
neuroimaging studies favor setting $\lambda_1, \lambda_2$ globally,
but we find
improved performance by varying them across voxels (Section~\ref
{secresults}), as
chosen by cross-validation [implemented in the \texttt{glmnet} MATLAB package
by \citet{friedmanglmnet}].

As with ridge regression, the elastic net estimate can be viewed as the MAP
estimate of a Bayesian model. As shown by \citet{kyung2010}, the required
prior is a gamma-scale mixture of Gaussians:
%e3 #&#
%
\begin{eqnarray}
\label{eqnbayeselasticnet} y_{vt} \mid\mu_v, \mathbf{x}_t,
{\bolds{\beta}}_v, \sigma_v^2 &\sim& N\bigl(
\mu_v + \mathbf{x}^\top_t {\bolds{
\beta}}_v, \sigma_v^2\bigr),\nonumber
\\
{\bolds{\beta}}_v \mid\sigma_v^2,
\mathbf{D}_{\tau}^{*} &\sim& N\bigl(0, \sigma_v^2
\mathbf{D}_{\tau}^{*} \bigr),
\\
\tau_1^2,\ldots,\tau_P^2 &\sim&
\prod_{j=1}^P \frac{\lambda_1^2}{2}
e^{-\lambda_1^2\tau_j^2/2} d \tau_j^2,\qquad \tau_1^2,
\ldots,\tau _P^2 > 0,
\nonumber
\end{eqnarray}
where $\mathbf{D}_{\tau}^{*} = \operatorname{Diag}\{(\tau_i^{-2} +
\lambda_2)^{-1}\}$ for all
$i$.

%s2.3 #&#
\subsection{Hierarchical Bayesian small area model}
\label{secSAE}

It is biologically plausible that voxels within the same ROI respond similarly
to stimuli. Penalization methods, such as the elastic net, make
estimates of
regression coefficients more precise via stabilization but do not pool
information from related voxels. In contrast, techniques for stabilizing
parameter estimates by partially pooling information across, or borrowing
strength from, related areas have been extensively developed in the literature
on small area estimation (henceforth SAE) [\citet{rao2003}]. While not
traditional in
neuroscience, SAE is well known to be effective at shrinkage when there are
multiple regions [\citet{pfeffermann2013}], here ROIs. Hence, we
explore simple
SAE methods for regularization which incorporate ROI-level effects, without
completely pooling within ROIs.

The SAE literature typically accomplishes partial pooling using hierarchical
Bayesian (HB) models, so we follow that precedent. As before, we model the
activity ${y}_{vt}$ in a voxel $v$ as a linear combination of the stimulus
features~$\mathbf{x}_t$:
%e4 #&#
%
\begin{eqnarray}
\label{eqnunitlevel} y_{vt} &=&\mathbf{x}_t^\top(
\mathbf{z}_v + \mathbf{u}_{A(v)}) + \varepsilon_{vt}
\nonumber\\[-8pt]\\[-8pt]\nonumber
&=& \mathbf{x}_t^\top{\bolds{
\beta}}^{SA}_{v} + \varepsilon_{vt},
\end{eqnarray}
where $A(v)$ is the ROI containing voxel $v$, $\mathbf{u}_{a}$ is a coefficient
vector common to all voxels in area $a$, and $\mathbf{z}_v$ is the coefficient
vector specific to voxel $v$. We have
\begin{eqnarray*}
% \label{eqnsaemodel}
{ y_{vt}} \mid{\bolds{\beta}}_v^{SA}, \sigma^2_v&\sim&\mathcal{N}\bigl( \mathbf{x}_t^\top{
\bolds{\beta}}^{SA}_{v},\sigma^2_v
\bigr),
\\
\bolds{\beta}^{SA}_v &=& \mathbf{u}_{A(v)}+
\mathbf{z}_v,
\\
\mathbf{z}_v \mid\nu^2_v&=& \mathcal{N}
\bigl(0,\nu^2_v\mathbf{I}\bigr),
\\
\mathbf{u}_a \mid\alpha^2_a&\sim&
\mathcal{N}\bigl(0,\alpha^2_a\mathbf {I}\bigr),
\\
\sigma^2_v&\sim&\mathcal{IG}(a,b),
\\
\alpha^2_a&\sim&\mathcal{IG}(c,d),
\\
\nu^2_v&\sim&\mathcal{IG}(e,f),
\end{eqnarray*}
where $a,b,c,d,e$, and $f$ are user-fixed hyperparameters, and
$\mathcal{IG}$ $(\mathrm{shape},\mathrm{scale})$ is the inverse gamma
distribution.
The full conditional distributions of all parameters are straightforward
(Appendix~A of the Supplementary Article [\citet{suppA}]), so the model
can be estimated
effectively using partially parallelized Gibbs sampling.

Just as ridge and the elastic net have Bayesian interpretations, the MAP
estimates of this Bayesian SAE model can be seen as a penalized least-squares
estimate. Such an estimate is (surprisingly) close to the estimate delivered
by ridge regression, for the following reason: the SAE model has a Gaussian
prior distribution $\mathbf{z}_v|\nu^2_v \sim\mathcal{N}(0,\nu
_v^2\mathbf{I})$
for the regression coefficients specific to voxel $v$, and the voxel-specific
variance has an inverse gamma prior distribution, where $\nu^2_v \sim
\mathcal{IG}(e,f)$. Due to this, the {marginal} prior distribution of
$\mathbf{z}_v$ is a scaled $t$-distribution, which is well
approximated by a
Gaussian for reasonable values of the hyperparameters (see
Appendix~B  of the Supplementary Article [\citet{suppA}] for
details). Section~\ref{ridgesaerelation}
revisits the statistical implication of this mathematical
approximation, which
is that the posterior mode of the HB model must actually be close to
the ridge
regression estimate.

%s2.4 #&#
\subsection{Spatial smoothing}

\label{secsmoothing}

Neuroimaging data is extremely noisy, and estimates have high
variance, even after shrinkage. Much of this noise occurs at high spatial
frequencies [\citet{ashby2011}, Chapter~4], and spatial smoothing
can help
reduce the variance. Since nearby voxels often tend to share activation
patterns, spatial averaging may cancel out such noise but maintain signal.
Biologically, nearby voxels should tend to respond similarly to
stimuli, since
recordings of individual cells show that many areas of the brain have a regular
spatial organization in their responses to stimuli [\citet{shepherd1994}].
While the length scales over which individual neurons' responses vary
do not
coincide with the sizes of voxels, which generally contain many cells with
heterogenous properties, it is still the case that nearby voxels should have
correlated responses to stimuli. Since the noise in fMRI data is often
at much
higher spatial frequencies than the signal from voxels, it is
reasonable to
think that spatially smoothing the activity will enhance the signal-to-noise
ratio. This is often done as a preprocessing step [\citet
{ashburner2008spm8}],
but we examine it here as a means of stabilizing parameter estimates.

We explore two kinds of spatial smoothing: \emph{nearest-neighbor voxel-level}
and \emph{ROI area-level smoothing}. First we introduce these
two forms of smoothing, and then consider smoothed OLS estimates.

%s2.4.1 #&#
\subsubsection{Nearest-neighbor voxel-level and ROI area-level smoothing}

\emph{Nearest-neighbor voxel-level smoothing} replaces every voxel by the
local average of its nearby voxels. This is done either for the activity
levels $\by_{v}$ or the parameter estimates ${\bolds{\beta}}_v$.
Lacking more
anatomically-based metrics, we define ``nearness'' using standard $\ell_p$
distances of two vectors $\mathbf{r}_1$ and $\mathbf{r}_2$:
\[
\llVert \mathbf{r}_1 - \mathbf{r}_2 \rrVert
_p \equiv{ \bigl( \llvert r_{11}-r_{21}\rrvert
^p + \llvert r_{12}-r_{22}\rrvert ^p
+ \llvert r_{13}-r_{23}\rrvert ^p
\bigr)}^{1/p}.
\]
When $p=2$, this is Euclidean distance and the $\ell_p$ ball around a voxel
contains all other voxels whose centers fall within the given radius. However,
when $p=1$, the $\ell_p$ ball is a tetrahedral pyramid. We choose a smoothing
range or radius separately for each voxel by cross-validation, and
replace its
value by the average over all voxels within the $\ell_p$
ball.\footnote{For a
given radius, the $\ell_1$ ball contains fewer voxels than the $\ell
_2$, and
both are smaller than the $\ell_{\infty}$ ball. The latter did so
poorly in
trials that we only consider $\ell_1$ and $\ell_2$.}

\emph{ROI area-level smoothing} is defined through solving an
optimization problem. Taking the set of regression coefficients in one
ROI A,
$\bBeta_A:= \{{\bolds{\beta}}_v\}_{v \in A}$, which is a $P \times
|A|$ matrix. We
penalize large differences between regression coefficients of voxels in the
same area. In the Bayesian setting,\vspace*{2pt} these are the voxel-wise Bayes estimates.
Specifically, for each ROI $A$, define $\tilde{\bBeta}_A$ as
\[
\tilde{\bBeta}_A= \mathop{\argmin}_{\tilde{\bBeta} = \{\tbb_v\}_{v \in
A}} \qquad\sum
_{v \in A} \llVert \tbb_v - {\bolds{
\beta}}_v\rrVert _2^2 + \gamma\sum
_{i,j \in A} q^A_{ij} \llVert
\tbb_i - \tbb_j\rrVert _2^2,
\]
with penalty factor $\gamma$ and $|A| \times|A|$ similarity matrix
$\mathbf{Q^A}$. Fixing $q^A_{ij} =1$ for all
$i,j \in A$, leads to more uniform smoothing. However, letting
\[
q^A_{ij} = \exp\bigl\{-d(i,j)^2/h^2
\bigr\}
\]
if $i,j \in A$, where $d$ is the Euclidean distance
between the locations of voxels $i$ and $j$ and $h$ is a bandwidth,
allows closer voxels to be more
influential. Since the above optimization problem splits across the
dimensions of ${\bolds{\beta}}_v$, we get $P$ independent
optimization problems. Denoting the $p$th row of $\tilde{\bBeta_A}$
as $\tbb^A_{p}$, we find
\[
\sum_{i,j \in A} q^A_{ij} (
\tbb_{ip} - \tbb_{jp})^2 = \tbb_{p}^{A\top}
\bolds{\Omega}_A \tbb^A_{p},
\]
where $\bolds{\Omega}_A:= 2(D^A - Q^A)$ is twice the graph Laplacian
formed using $Q^A$ as the adjacency matrix and $D^A$ as a diagonal
matrix whose $i$th entry is $\sum_{j} Q^A_{ij}$
[\citet{luxburg2007}, Proposition~1].
Hence,
$
\tilde{\bBeta}_A = (I + \gamma\bolds{\Omega}_A)^{-1}\bBeta_A$.
Parameters $\gamma$ and $h$ are chosen by
cross-validation.

%\textcolor{red}{ Noticing that $\Omega$ acts
%like a graph Laplacian allows us to use computational tricks to speed
%up the
%inversion (citation from AR) => LW: I removed this sentence because if
%$\bb$ contains only the voxels 1...V in ROI A, then $\Omega$ is $V
%\times V$ and we invert $\tilde{\bb} = (I + \gamma\Omega)^{-1}\bb$
%without a computational trick. However, if we use $\bb=:\{\beta_v\}
%all  v $, then we can say the following sentence: ``
%Noticing that $\bolds{\Omega}$ is a block diagonal matrix allows us to
%speed up the inversion by computing the smoothed estimates $\tilde{
%\bBeta}_A:= \{\tilde{\bbeta}_v\}_{v \in A}$ for each ROI $A$
%individually as: $
%\tilde{\bBeta}_A ~=~ (I + \gamma\bolds{\Omega}_A)^{-1}\bBeta_A$,
%where ${\bBeta}_A:= \{{\bbeta}_v\}_{v \in A}$.

%\textcolor{red}{Note that in Appendix E $\bb$ could mean both things.}

%s2.4.2 #&#
\subsubsection{Smoothed OLS}

Since OLS estimates are linear in $y_{vt}$ and covariates are identical across
voxels, smoothing ${\bolds{\beta}}_v$ is equivalent to smoothing
$y_{vt}$. At any voxel
$v$, let $\mathcal{S}_v$ be the set of voxels which are combined with
it in
smoothing, with the weight of voxel $u \in\mathcal{S}_v$ in the
smoothing for
$v$ being $c_{uv}$. These weights are functions of the radius of
smoothing in the nearest neighbor version, or of $\gamma$ and $q$ for ROI-level
smoothing. Then the smoothed estimate at $v$ is
\begin{eqnarray*}
\hat{\bar{\bolds{\beta}}}_v & = & \sum
_{u \in\mathcal
{S}_v}{c_{uv} \hat{{\bolds{\beta}}}_u}
\\
& = & \sum_{u \in\mathcal{S}_v}{c_{uv} \bigl(
\X^\top\X\bigr)^{-1} \X^\top \by_u}
\\
& = & \bigl(\X^\top\X\bigr)^{-1} \X^\top\sum
_{u \in\mathcal{S}_v}{c_{uv} \by_u}
\\
& = & \bigl(\X^\top\X\bigr)^{-1} \X^\top\bar{
\by}_v,
\end{eqnarray*}
which is the OLS estimate with the smoothed response $\bar
{y}$.\footnote{We
are certainly not the first to note that linear smoothing commutes with OLS
estimation---see, for example, \citeauthor{Friston-et-al-against-localizers} [(\citeyear{Friston-et-al-against-localizers}), page~12].}

Despite the simplicity of the technique, smoothed OLS produces results
quite comparable to regularization methods such as ridge regression (see
Section~\ref{secresults}).

The equivalence of smoothing parameter estimates and smoothing the activity
does not hold with our other, nonlinear estimators. When we report results
for combinations of smoothing with other forms of regularization, we are
smoothing the parameter estimates.

%s2.5 #&#
\subsection{Evaluation criteria}
\label{secevaluation}

Typically, cognitive neuroscientists engage in two forms of predictive inference
with fMRI: {\em forward inference}, from stimuli to configurations of activity
over the brain, and {\em reverse inference}, from patterns of activity to
stimuli. While these are often approached as two separate tasks with two
distinct sets of models, we perform both forward and reverse
inference, using a common model.

Forward inference is a regression problem, where the regression models
reviewed above can be applied immediately. Our evaluation criterion for
forward inference is the voxel-wise residual sum of squares, normalized
by the
total sum of squares.

Reverse inference is more delicate. If we were primarily interested in
decoding stimuli from observed neural activity, we could follow the usual
practice in fMRI data analysis of estimating ``tailored'' classifiers or
discriminative models [\citet
{poldrack2008,pereira2009,yarkoni2011}]. These
might be accurate for the particular conditions they were trained on,
but by
construction they cannot generalize to previously unseen stimuli,
unless they
predict as an intermediate step the individual features of the stimuli
and then
identify the correct stimuli based on the decoded features
[\citet{sudre2012tracking}]. Moreover, discriminative models do
not directly
represent anything about how the brain processes information, which is
the main
point of scientific interest.\footnote{Symbolically, scientists want
to know
about $p(Y|X)$, while discriminative models at best give $p(X|Y)$,
which, by
Bayes's rule, combines $p(Y|X)$ and the distribution of stimuli
$p(X)$.} As
shown by \citet{haufe2014interpretation}, the parameters learned
in a decoding
model, corresponding to each voxel's contribution in a decoding task,
cannot be
readily used to infer if a voxel is representing a task of interest. For
example, some voxels that represent a background process unrelated to
the task
might receive a high regression weight that serves to subtract that process
from the voxels that are informative to the
task.\footnote{\citet{haufe2014interpretation} do suggest a
method to enable a
neurophysiological interpretation of the parameters of linear decoding
models.}

As shown by \citet{mitchell2008predicting}, it is possible to use
a forward
model to do reverse inference, and doing so provides an additional
check on the
forward model's ability to represent how the brain processes stimuli.
This is
an instance of ``zero-shot classification'' [\citet
{palatucci2009zero}] adapted
to the neural prediction task, where the model is trained as usual, but with
the data for some stimuli held out. The trained model is faced with the
$y_{vt}$ for a held-out stimulus condition in a particular voxel $v$,
and the
two sets of features for the correct stimulus condition and another unseen
stimulus condition chosen at random. Next, the trained model makes a
prediction for both stimuli, and the data point is assigned to the stimulus
whose predicted activity is closer to the observed $y_{vt}$. By design, chance
performance for the balanced binary reverse-inference task is 50\%.

This is easily extended to from one voxel $v$ to the entire brain. We compute
the distance between the observed $y$ and the predictions of the
forward model
as a weighted sum of the voxel-wise distances, where the weights depend
on the
classification accuracy of the individual voxels on the training set. Each
voxel is weighed by the inverse of its rank when the per-voxel classification
accuracies are sorted in decreasing order. Now, each stimulus is not
represented by the original features, but instead by the weighted error
in its
forward model's predictions of neural activity.\footnote{This is
analogous to the
way support vector machines and other kernel classifiers expand the dimension
of the feature space by computing many nonlinear functions of the features
[\citet{Cristianini-Shawe-Taylor}], and to the use of generative model
likelihoods to define discriminative kernels
[\citet{Jaakkola-Haussler-on-Fisher-kernels}].} (As before, voxel-wise
accuracies are determined through cross-validation.) One consequence of the
weighting scheme is that whole-brain reverse inference is highly
accurate if
there are only a few high-accuracy voxels. Of course, whole-brain
classification can also be accurate even if no one voxel has high accuracy.

%pa2.5.subsubsection.1 #&#
\subsubsection*{Validation sets and cross-validation}

We evaluate both forward and reverse inferences with nested 10-fold
cross-validation. 10\% of the data is held for testing. We then use the
remaining training set (90\%) to compute the different estimates. For
\textup{E2}, we throw out 5 images on the boundaries of the training set and the test
set to insure that there is no signal leakage from the training to the test
set due to the slow decay of hemodynamic responses, causing unintended
correlations. If we choose not to smooth the estimates, then we proceed as
follows with the training set.

For ridge regression, we use generalized cross-validation
[\citet{golub1979generalized}] to approximate leave-one-out
cross-validation
error for different $\lambda_v$ values at each voxel. For the elastic
net, we
use the ten-fold cross-validation option provided in \citet
{friedmanglmnet} to
chose the regularization parameters. Finally, for SAE, the level of
regularization is determined by the posterior mean variance of ${\mathbf z}_v$,
since high variance corresponds to the model being able to choose the parameter
freely, that is, low regularization. The posterior mean variance of
${\mathbf z}_v$
is determined automatically by the Gibbs sampler.

If we choose to smooth the estimates, then in order to pick the smoothing
parameter for every voxel and every estimator, we run a nested cross-validation
loop. That is, within the 90\% training portion of the data, 80\% is
randomly selected
as ``inner-fold training'' data, and 10\% is randomly selected as a
validation set. The inner-fold
training is done exactly as in the previous paragraph. Smoothing parameters
are then set using the average single-voxel classification accuracy on the
validation set.

After training, the out-of-sample performance of both unsmoothed and smoot\-hed
estimators is reported using the testing set.
Thus, the parameters never adapt to the testing set, and we report valid
estimates of out-of-sample performance.

%s3 #&#
\section{Results}
\label{secresults}

Our main findings are as follows: using cross-validation to pick tuning
parameters
separately for each voxel:
\begin{longlist}[2.]
\item[1.] Regularization offers small but real gains in forward prediction;
\item[2.] Regularization does not seem to offer improvement in reverse prediction
at the individual voxel level, or whole-brain reverse inference;
\item[3.] All forms of regularization work about equally well for prediction;
\item[4.] Regularization succeeds in making parameter estimates more precise;
\item[5.] The spatial pattern of regularization is highly informative:
voxels where
unregularized OLS is least accurate are precisely the ones which are more
heavily smoothed or regularized under cross-validation.
\end{longlist}
We explain these points in turn.

%s3.1 #&#
\subsection{Prediction}
\label{secaccuracy}

To summarize, while all models and methods had some predictive ability
in both
experiments, none of them clearly dominated the others. Model checking,
discussed in Appendix~C of the Supplementary Article [\citet{suppA}],
shows that this was not because
the models were grossly inappropriate, though they are somewhat misspecified.

%pa3.1.subsubsection.1 #&#
\subsubsection*{Forward inference}

%f1 #&#
%
\begin{figure}\vspace*{6pt}%[t]

\includegraphics{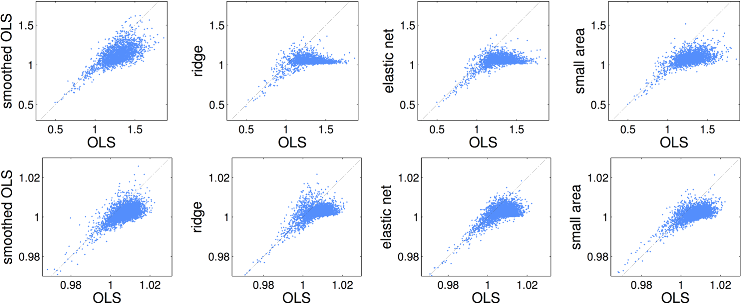}

\caption{Effect of regularization on out-of-sample normalized RSS
($\mathrm{RSS}/\sigma^2$) for \textup{E1} (top) and \textup{E2} (bottom). For each of the plots, the
OLS $\mathrm{RSS}/\sigma^2$ (horizontal axis) is contrasted with the modified
$\mathrm{RSS}/\sigma^2$ after OLS smoothing for ridge, elastic net, or small area
shrinkage (vertical axis). For both experiments, the four methods
result in
smaller $\mathrm{RSS}/\sigma^2$ on average. Furthermore, for all the methods, the
predicted activity in the bad voxels (i.e., voxels where $\mathrm{RSS}/\sigma
^2$ is
larger than 1) is pushed toward zero. This is visible by the
$\mathrm{RSS}/\sigma^2$
values being reduced toward 1. In other words, shrinkage and smoothing are
forcing the estimated parameters to be almost zero if the voxel is noisy
and there is nothing that can be predicted. Note that the scales of the axes
are different for the two experiments.}
\label{figrssbeforeandaftershrinkage}
\end{figure}

%f2 #&#
%
\begin{figure}%[b]

\includegraphics{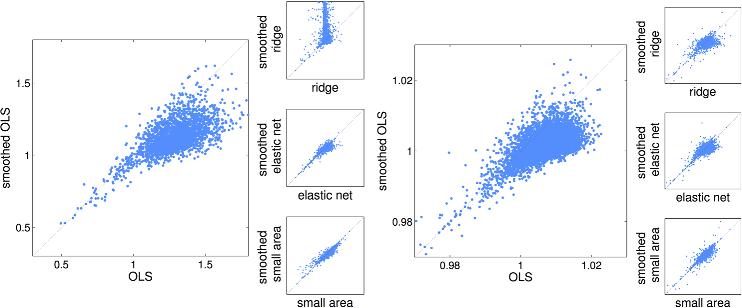}

\caption{Normalized RSS for unsmoothed and smoothed estimators for \textup{E1} (left)
and \textup{E2} (right). The larger panels show voxel-wise normalized residuals
($\mathrm{RSS}/\sigma^2$) for OLS before smoothing (horizontal axis) and after
(vertical), showing the value of spatial smoothing for forward inference.
The smaller panels consist of the same comparison for ridge regression
(top), the elastic net (middle), and the small area model (bottom), showing
that combining smoothing and shrinkage is if anything worse than shrinkage
alone. The axes for the smaller panels have been omitted for clarity: they
correspond to the larger panels axes.}\vspace*{6pt}
\label{figrssbeforeandaftersmoothing}
\end{figure}

All our methods had nontrivial ability to do forward prediction in both
experiments for some of the voxels, which should be the voxels that are
implicated in visual processing (Figure~\ref
{figrssbeforeandaftershrinkage}). All methods of regularizing OLS,
including spatial smoothing, led to generally small but significant
improvements. The improvement is seen in the noisy voxels: the high RSS in
those voxels is greatly reduced when shrinkage or smoothing is used,
effectively driving the prediction in the noisy voxels to zero. Since
results for both neighborhood- and ROI-based smoothing were nearly
identical, we report only those for smoothing over $\ell_2$ balls (however,
see Section~\ref{secdiscussion}.) Combining smoothing with shrinkage did not
help forward inference; if anything, it often made it worse than either alone
(Figure~\ref{figrssbeforeandaftersmoothing}).

%All our methods had non-trivial ability to do forward prediction in
%both
%experiments (Figures \ref{figrsse1}, \ref{figrsse2}). All methods of
%regularizing OLS, including spatial smoothing, led to generally small
%but
%significant improvements.\footnote{Since results for both
%neighborhood- and
% ROI- based smoothing were nearly identical, we report only those for
% smoothing over $\ell_2$ balls. However, see Section~\ref{secdiscussion}.}
%Combining smoothing with shrinkage did not help forward inference; if
%anything
%it often made it worse than either alone (Figure
%\ref{figrssbeforeandaftersmoothing}).

%\begin{figure}
% \begin{center}
% \textsc{Insert something like Figure \ref{figaccuracye1}, but for RSS
%in \textup{E1}}
% \end{center}
% \caption{Voxel-wise normalized RSS in experiment \textup{E1}, averaging over
%voxels
% and subjects, for all combinations of estimators and smoothing.}
% \label{figrsse1}
%\end{figure}
%
%\begin{figure}
% \begin{center}
% \textsc{Insert something like Figure \ref{figaccuracye1}, but for RSS
%in \textup{E2}}
% \end{center}
% \caption{Voxel-wise normalized RSS in experiment \textup{E2}, averaging over
%voxels
% and subjects, for all combinations of estimators and smoothing.}
% \label{figrsse2}
%\end{figure}

%pa3.1.subsubsection.2 #&#
\subsubsection*{Reverse inference}

The effect of regularization on single-voxel reverse inference is ambiguous
(see Appendix~E  of the Supplementary Article [\citet{suppA}]):
accuracy goes up in some voxels and
down in others, with no change over all. The classification accuracy of the
good voxels varies much less across the different estimators than the accuracy
of the bad voxels (see figures in Appendix~E of the Supplementary Article [\citet{suppA}]).

\begin{figure}%[h!]

\includegraphics{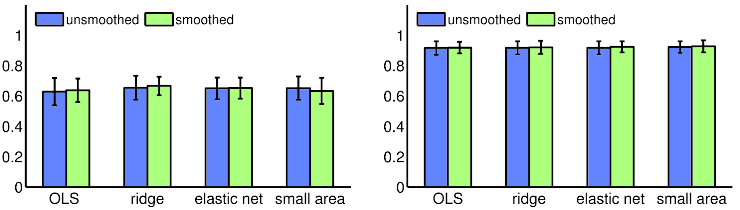}

\caption{Whole-brain classification accuracy in experiments \textup{E1} (left)
and \textup{E2} (right), averaging over
subjects, for all combinations of estimators and smoothing.
Regularization choice or the presence or absence of smoothing does not
affect whole-brain classification accuracy.}
\label{figaccuracye12}
\end{figure}

Turning to whole-brain reverse inference, all methods, with and without
smoothing, did much better than the chance rate of 50\% in both experiments
(Figure~\ref{figaccuracye12}). However, the differences between
methods are
negligible, and certainly smaller than the fold-to-fold variability of
cross-validation. This includes unregularized OLS.

%f3 #&#
%

%\begin{figure}%[h!]
% \begin{center}
% \end{center}
% \caption{Whole-brain classification accuracy in experiment \textup{E2},
%averaging over
% subjects, for all combinations of estimators and smoothing.}
% \label{figaccuracye2}
%\end{figure}

All our methods predict equally well (up to experimental
precision), which is surprising. We can rationalize
the elastic net performing
about as well as ridge regression on the grounds that the former
extends the
latter by adding an $L_1$ penalty, which might be unnecessary. Ridge
regression is also linked to our hierarchical small area model via an
approximation result (Appendix~B  of the
Supplementary Article [\citet{suppA}]). However, such connections
do not account for why all three forms of shrinkage perform about the
same as smoothed OLS or unsmoothed OLS.

We do find a partial explanation from the way we do whole-brain
classification (Section
\ref{secevaluation}). Recall that we classify a pattern of activity as
belonging to the
stimulus whose predicted activity pattern is closest, but weight each
voxel in
this distance calculation depending on its individual classification
accuracy. Thus,
the weights are often dominated by a fairly small number of
highly discriminative voxels. These voxels tend to also be ones
where the forward model fits well, and cross-validation or Gibbs
sampling selects little or no
regularization for them. To support these claims, we
examine the effects of regularization on the parameter estimates and the
spatial patterns of regularization.

%s3.2 #&#
\subsection{Regularization}

%s3.2.1 #&#
\subsubsection{Evidence of successful regularization}

In light of the surprising predictive equivalence of our different
methods with
each other and with OLS, it is worth verifying that our regularlizers
were in
fact regularizing the estimation. From the standpoint of small area estimation
theory, the crucial question is whether the parameter estimates are more
precise than the ``direct'' estimates of OLS. That is, do the new
estimates show smaller standard errors, or smaller coefficients of variation,
than the direct estimates?

\begin{figure}%[h!]

\includegraphics{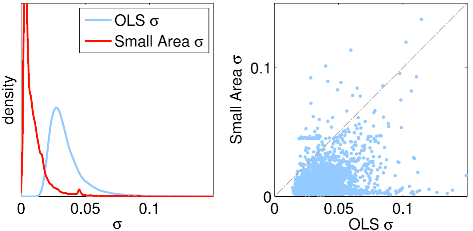}

\caption{Left: Histograms of one regression coefficient's standard
errors in
\textup{E1}, aggregating over all voxels, for both OLS and SAE. The sharp peaking
of the latter histogram, to the left of the former, indicates that the
typical parameter estimate has been made much more precise by the
hierarchical model. Right: scatter-plot of the same standard errors. Most
of the points fall below the diagonal, so most parameters are being
estimated more precisely. Other coefficients and methods of regularization
behaved similarly.}\label{figissmall}
\end{figure}

%While we omit exhaustive demonstration of this point,
Results like in Figure~\ref{figissmall} are typical across the
coefficients and the regularizers.
After regularization, most parameter estimates for most voxels had
significantly smaller standard errors, sometimes much smaller. This was true
even while using cross-validation to pick how much to regularize each voxel.
(See Appendix D of the Supplementary Article [\citet{suppA}] for
additional documentation.)

%f4 #&#
%

%f5 #&#
%
\begin{figure}

\includegraphics{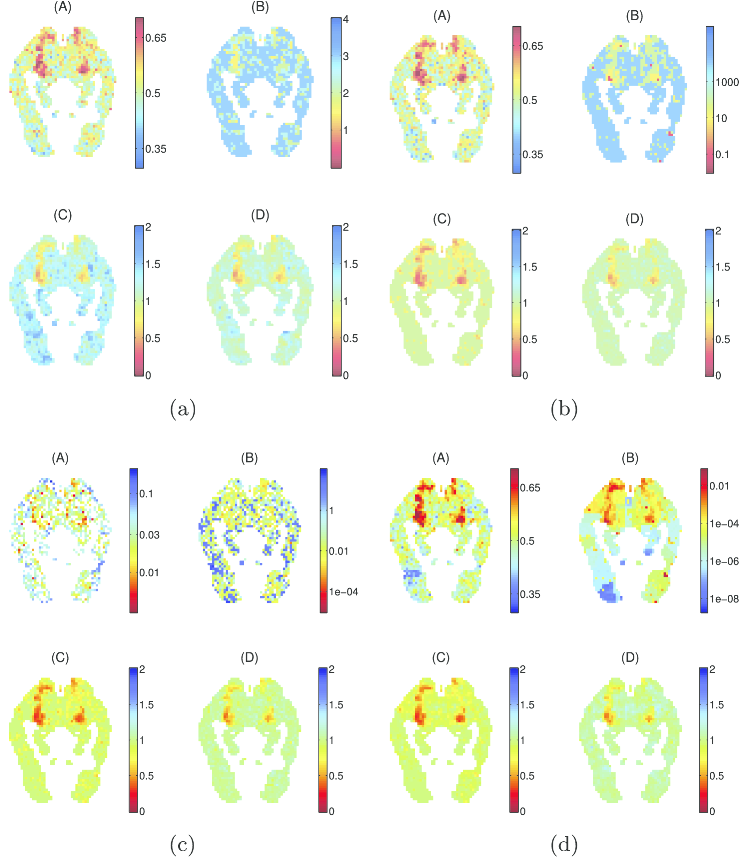}

\caption{Voxel-wise results for each method along one horizontal
brain slice in experiment \textup{E1}. Color schemes are flipped so that red
always represents ``good'' and blue, ``bad.'' Note the similar patterns
of classification accuracy in plots \textup{(a)-(A)}, \textup{(b)-(A)}, and \textup{(d)-(A)}. Also note how
predictive performance [subfigures \textup{(A)} and \textup{(D)}] is inversely related to the
degree of regularization in every case [the smoothing
radius for OLS \textup{(a)}, the $\lambda$ penalty for ridge \textup{(b)}, the $\lambda_1$
and $\lambda_2$ penalties for the elastic net \textup{(c)}, or the posterior
mean variance in the small area
model \textup{(d)}---high variance means low regularization]. Finally, see that
in many cases the in- and out-of-sample errors for ``good'' voxels are nearly
the same.
\textup{(a)} OLS:
\textup{(A)}~classification accuracy;
\textup{(B)}~smoothing radius;
\textup{(C)}, \textup{(D)}~normalized out-of-sample RSS pre- and post-smoothing.
\textup{(b)}~Ridge:
\textup{(A)}~classification accuracy;
\textup{(B)}~$\lambda$ parameter;
\textup{(C)}, \textup{(D)}~normalized RSS in- and out-of-sample.
\textup{(c)}~Elastic Net:
\textup{(A)}~$\lambda_1$ (lasso penalty);
\textup{(B)}~$\lambda_2$ (ridge penalty);
\textup{(C)}, \textup{(D)}~normalized RSS in- and out-of-sample.
\textup{(D)}~Small Area:
\textup{(A)}~classification accuracy;
\textup{(B)}~posterior mean variance of ${\mathbf z}_v$;
\textup{(C)}, \textup{(D)}~normalized RSS in- and out-of-sample.}
\label{figbrain} \label{figOLS} \label{figridge}\label{figgibbs}
\end{figure}

%f6 #&#
%
\begin{figure}

\includegraphics{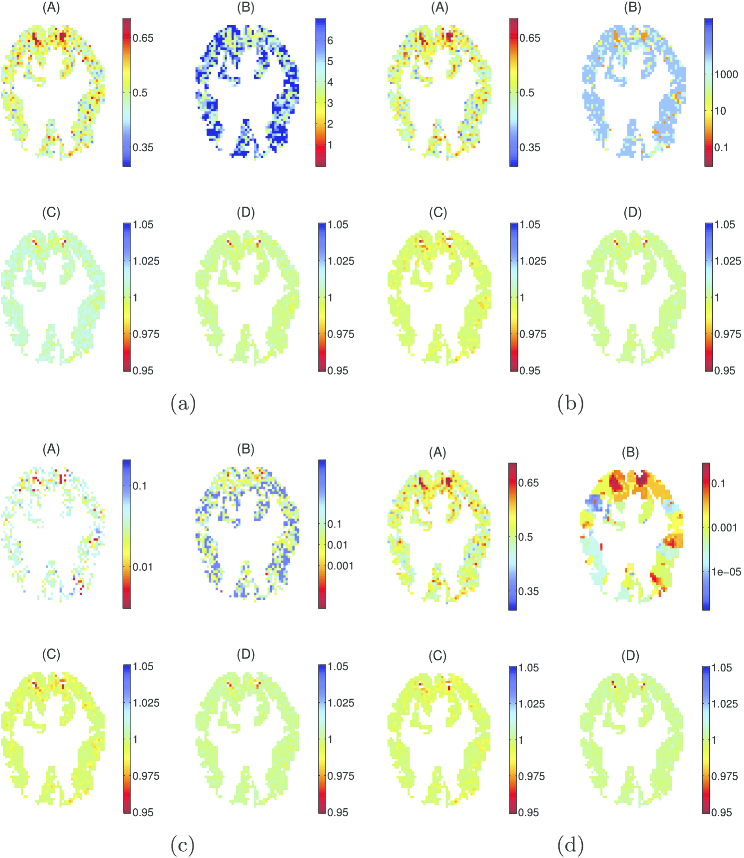}

\caption{Voxel-wise results for each method along one horizontal
brain slice for experiment \textup{E2}. Color schemes are flipped so that red
always represents ``good'' and blue, ``bad.'' See Figure~\protect\ref
{figbrain}
for more details. As in Figure~\protect\ref{figbrain}, predictive performance
[subfigures \textup{(A)} and \textup{(D)}] is inversely related to the degree of
regularization in every case.
\textup{(a)} OLS:
\textup{(A)}~classification accuracy;
\textup{(B)}~smoothing radius;
\textup{(C)}, \textup{(D)}~normalized out-of-sample RSS pre- and post-smoothing.
\textup{(b)}~Ridge:
\textup{(A)}~classification accuracy;
\textup{(B)}~$\lambda$ parameter;
\textup{(C)}, \textup{(D)}~normalized RSS in- and out-of-sample.
\textup{(c)}~Elastic Net:
\textup{(A)}~$\lambda_1$ (lasso penalty);
\textup{(B)}~$\lambda_2$ (ridge penalty);
\textup{(C)}, \textup{(D)}~normalized RSS in- and out-of-sample.
\textup{(D)}~Small Area:
\textup{(A)}~classification accuracy;
\textup{(B)}~posterior mean variance of ${\mathbf z}_v$;
\textup{(C)}, \textup{(D)}~normalized RSS in- and out-of-sample.}
\label{figbrainns}
\end{figure}

%s3.2.2 #&#
\subsubsection{Spatial patterns of regularization and their implications}
\label{secpatterns}

The\break strength of regularization chosen by cross-validation is not
uniform or
even random across the brain. It shows quite pronounced,
and informative, spatial structure, closely connected to how well voxels
predict without regularization.

Figures~\ref{figbrain} and \ref{figbrainns} depict the relationship between
the degree of regularization imposed by our methods, and several
measures of
predictive accuracy. For two horizontal slices of the brain, these figures
illustrate how classification accuracy varies, how strongly regularized each
voxel is, and how well the regression model does in and out of sample. (The
accuracy plot is omitted for the elastic net to show both penalty factors.)
Appendix F of the Supplementary Article [\citet{suppA}] provides the
corresponding plots for the
entire brain. The plots provided are for two subjects, one from each
experiment. The other subjects present a very similar pattern of correspondence
between voxels with high performance and weak regularization.

As Figure~\ref{figOLS}(a)--(d) shows, there is an inverse
relationship between predictive performance [subfigures (A) and (D)] and the
degree of regularization [subfigures~(B)] that was chosen by
cross-validation, whether that is the smoothing
radius for OLS (a), the $\lambda$ penalty for ridge (b), the $\lambda_1$
and $\lambda_2$ penalties for the elastic net (c), or the small area
model (d), where low regularization corresponds to a high variance
parameter, that is, good voxels are allowed to pick their parameters freely.
For the elastic net, good voxels have more lasso-like penalties, as they
are voxels sensitive to {\em some} of the stimulus features. Smoothing
acts as a regularizer for OLS, as seen by the reduced prediction in the
bad voxels from subfigure \mbox{(a)-(C)} to subfigure (a)-(D).
Thus, voxels with stronger signals (as reflected by higher
accuracy) needed less regularization. Voxels with high accuracy
[Figure~\ref{figOLS}(a), (b) and (d), part A] and especially
voxels with low prediction error [subfigures (D)] are sparse and spatially
clustered. Other voxels are by comparison noisy and more heavily regularized
[subfigures (B)].

The correspondence between good classification accuracy and weak regularization
explains the single voxel accuracy results mentioned in Section~\ref
{secaccuracy}
and in Appendix E of the Supplementary Article [\citet{suppA}]. In
good voxels, classification
accuracy is not significantly affected by regularization since the penalty
parameter is weak. In the bad voxels, the strong regularization forces the
model to learn near-zero weights, and the leftover noise has a
``random'' effect
on the single voxel classification accuracy, sometimes resulting in slight
improvement, and sometimes in slight decrease.

For \textup{E1}, the predictive voxels are clustered in the occipital cortex,
which is
well known to be heavily involved in visual processing [\citet
{shepherd1994}].
For \textup{E2}, the predictive voxels involve a smaller part of the occipital cortex,
as well as some small clusters of voxels in more anterior regions associated
with language comprehension (such as the left temporal lobe).

%

%s4 #&#
\section{Discussion}
\label{secdiscussion}

%s4.1 #&#
\subsection{Ridge and SAE}
\label{ridgesaerelation}

We have shown that different forms of regularization predict about equally
well. Moreover, they give similar parameter estimates, especially the SAE
model of \eqref{eqnunitlevel} and ridge regression. As already
explained in
Section~\ref{secSAE}, the marginal prior distribution of $\beta_v$ is an
inverse-gamma variance mixture of Gaussians, which is a
$t$-distribution, where
$\beta_v \sim t$. With even a moderate number of degrees of freedom in
the $t$,
the marginal prior on $\beta_v$ is quite close to being Gaussian (Appendix~B
of the Supplementary Article [\citet{suppA}]). Similarly, the
marginal prior on $u_i$ is also
a $t$-distribution. Since $\beta_v$ and $u_{A(v)}$ are independent
{a priori}, the prior on $z_v$ is approximately Gaussian. Since the posterior
mode under a Gaussian prior matches ridge regression, the $z_v$
estimated from
\eqref{eqnunitlevel} will be close to the ridge regression
estimates. We have not been able to find this approximation result in
the literature, but suspect it is a rediscovery.

When we simulate from the SAE model, estimating that model shows better forward
prediction than OLS or even ridge regression (Appendix C.1 of the Supplementary Article [\citet{suppA}]).
The difference between SAE and ridge is small but systematic and significant.
However, when the surrogate data from the simulations is re-estimated with
erroneous assignments of voxels to ROIs, the advantage of the SAE model over
ridge regression vanishes. It {\em may} be that this is the way in
which the
SAE is misspecified, suggesting that a better choice of ROIs would lead to
superior prediction. However, we have not been able to rule out other possible
misspecifications.

%s4.2 #&#
\subsection{Computational costs}
\label{seccomputational}

While our four methods perform very similarly statistically, their
computational costs differ by orders of magnitude (Table~\ref{tableruntime}).
Smoothed OLS and ridge stand out as the most attractive methods, with ridge
pulling ahead due to its better behaved out-of-sample residuals.

%t1 #&#
%
\begin{table}[b]
\tabcolsep=0pt
\caption{Running times of the various procedures on the \textup{E1} data, using 8
Intel Xenon CPU E5-2660 0 cores (at 2.2~GHz), sharing 128~GB of RAM. Gibbs
sampling for the SAE model was parallelized over~the~cores}
\label{tableruntime}
\begin{tabular*}{\tablewidth}{@{\extracolsep{\fill}}@{}lccc@{}}
\hline
&\textbf{cpu time per fold} & \textbf{Clock time per fold} &\textbf{Total cpu time}\\
&\textbf{per subject}       & \textbf{per subject}         &\textbf{(with nested CV)}\\
\hline
OLS&$<$1~s&$<$1~s&$<$1~min\\
Ridge&55~s&4~s&7.5~h\\
Elastic net&3120~s&390~s&429~h\\
Small area&5540~s&740~s&762~h\\
Smoothing, nested CV&40~s&20~s&5.5~h\\
\hline
\end{tabular*}
\end{table}

Our simple and generic HB model is misspecified, not very firmly
grounded in
biology and, as Table~\ref{tableruntime} shows, computationally very costly.
With considerable attention to the biology, well-specified models and priors
might be crafted for specific applications, though at even greater
computational expense. Due to this, we do not advocate the Bayesian approach,
unless it could be combined with some way of {\em quickly} approximating
posterior distributions, for example, variational methods [\citet
{wainwright2008,broderick2013}] or consensus MCMC [\citet
{scott2013,xing2013}]. Such
extensions are beyond the scope of this paper.

%s4.3 #&#
\subsection{The detectability assumption}
\label{secspotlight}

As mentioned in Section~\ref
{secscientific-and-statistical-background}, it is
common in brain-imaging studies to do a separate regression for each
voxel on
the stimuli, and presume that only voxels with significant regression
coefficients are involved in processing the stimulus. This
assumption appears to have no neurobiological basis; we call it the
``detectability
assumption.'' In practice, neuroscientists recognize this leads to some number
of errors, both false positives and negatives. However, they often
presume that
these errors are random rather than systematic. Under the detectability
assumption, methods of regularization
%which shrink regression coefficients
%towards zero
might plausibly be seen as reducing the rate of false positives.
If true, this is an important advantage for regularized estimates, even
if they predicted no better than OLS. The spatial pattern of regularization,
under this assumption, is an indicator of which voxels are involved in
processing the stimulus features. This indication is strengthened by the
similarity of these patterns under different methods of regularization
(Section~\ref{secpatterns}).

Since constant-bandwidth smoothing the data spatially is a common fMRI
preprocessing step, usually followed by using OLS, it can be argued that
existing analyses are already doing some regularization. However, as
Figure~\ref{figrssbeforeandaftershrinkage} shows, our shrinkage methods
reduce prediction error in the noisy voxels somewhat more than does smoothing
OLS. Moreover, even if one preferred to use smoothed OLS rather than
shrinkage, we have shown that good voxels do not need to be regularized
as much
as noisy voxels. Therefore, the current approach can be improved by choosing
the smoothing parameter at every voxel.

Despite its ubiquity, it is hard to support the detectability assumption
neurobiologically. A crucial component of it is the presumption that the
hemodynamic response is systematically related to information processing.
While it is true that increased spiking rates within a voxel will lead
to a
hemodynamic response, animal experiments show that neural information
can be
encoded in the time intervals between spikes rather than the spiking rate
[\citet{Spikes-book}], and in the coordination of spiking across
neurons, which
may lie within the same voxel or be widely distributed across the brain
[\citet
{Abbott-Sejnowski-neural-codes,Ballard-Zhang-Rao-distributed-synchrony,Engel-Fries-Singer-synchrony,Fries-gamma-band}].
Further, many neural circuits work by inhibiting other neurons, and increasing
inhibition may either increase or decrease energy demands and so hemodynamic
responses, depending on fine-grained anatomical and physiological details
[\citet{logothetis2008}].

Such considerations undermine the link between changes in local spiking rates,
energy use by neurons, hemodynamic response, and actual neural
computation or
information-processing. If information is conveyed by timing, conveyed by
synchrony, distributed across large spatial volumes, or works through a balance
between excitation and inhibition, then much neural computation might be
invisible in the hemodynamic signal which fMRI measures. This leads to
systematic false negatives which are inevitable when working with fMRI.

Another difficulty with the detectability assumption is that it
presumes that
when a voxel's hemodynamic signal does respond to stimulus features, the
regression coefficients are always relatively large. Usually, ``relatively
large'' amounts to ``statistically significant.'' This all runs
together with the
absolute magnitude of the regression coefficients, the sample size (including
the duration of each experiment and the number of subjects), the
variance of
the stimulus features, and the extent to which the features are
correlated with
each other. With larger samples and higher-variance, less-correlated features,
smaller regression coefficients become significant. That is, there is more
power to detect small coefficients. The negative inference that certain voxels
are not involved in processing stimulus features presumes that
feature-sensitive voxels have coefficients large enough that the experiment
has substantial power to detect them.

Regularization does not necessarily improve this situation. While it does
avoid making a hard-and-fast decision based on significance, it is
still true
that the optimal amount of regularization generally declines with the sample
size. Moreover, small coefficients, being hard to estimate, could be heavily
penalized under cross-validation. Thus, while regularization may
reduce the number of false positives, this may be more than
counterbalanced by
an increase in false negatives, unless all nonzero coefficients are fairly
large.

In summary, the detectability assumption contains two parts---that neural
information-processing always shows up in the hemodynamic response, and that
associated regression coefficients are always either zero or large---which
our current knowledge of neurobiology does not support. Nonetheless,
without a
feasible replacement, we hesitate to reject outright an assumption
embraced by
so much of the neuroscientific community.

%s4.4 #&#
\subsection{Conclusion}
\label{secdisc}

Our main finding is that how we regularize, whether using shrinkage and or
smoothing, is much less important for prediction than regularizing
somehow (Sections
\ref{secaccuracy} and \ref{secpatterns}). All regularization methods
considered (ridge, elastic net, the small area HB model, and smoothed OLS)
improved forward and backward predictions about equally. When we
allowed the
degree of regularization to vary across the brain, voxels with strong signals
receive little regularization, while more noisy voxels are heavily regularized
[Figure~\ref{figOLS}(a)--(d)]. Furthermore, very similar patterns
emerged from all methods. Since the methods are similar predictively,
we favor
ridge and smoothed OLS on computational grounds. Ridge regression is already
widely used, but smoothed OLS should be added to the fMRI toolkit.

None of our methods were designed for fMRI problems and none were
informed by a
deep understanding of the physics of measuring hemodynamic response or
any type
of neuropsychological model. However, we hope that better predictions
can be
obtained through regularization methods that express neurologically relevant
forms of smoothness, sparsity, and similarity, rather than just being
``off the
shelf'' priors or penalties. We do not mean to be dogmatic about whether
neurobiological constraints should be expressed as objective functions
or as
stochastic processes, though we suspect that a penalized optimization approach
is more computationally tractable than a Bayesian approach. It is hard to
imagine a biologically sound Bayesian model leading to conjugate
priors. If a
Bayesian approach is taken, it should be biologically and
neurologically sound
and computationally efficient. Posterior approximation methods should
play a
crucial role, and we leave this for future exploration. Whether priors or
penalties, the regularizers of the future must be neural models.

%\begin{appendix}
%\section{}
%\end{appendix}

\section*{Acknowledgments}
We would like to thank the referees and the Associate Editor for comments that led
to major improvements of this paper.
The views in this paper are of the authors alone and not of the funding
agencies.

\begin{supplement}[id=suppA]
\sname{Supplementary Article}
\stitle{Appendix for ``Regularized brain reading with shrinkage and smoothing''}
\slink[doi]{10.1214/15-AOAS837SUPP} %[doi,text={...}] - jei reikia
%suskaldyti doi
\sdatatype{.pdf}
\sfilename{aoas837\_supp.pdf}
\sdescription{This supplement consists of six parts. It offers more
details about: (A) our Small Area model and Gibbs sampler, (B) the
Marginal Prior of the SAE Model, (C) model checking, (D) the effect of
regularization on variability, and (E) the effect of smoothing and
regularization on single voxel accuracy, as well as (F) whole brain
plots of the experimental results that are portrayed in Figures~\ref
{figbrain} and \ref{figbrainns} for a single slice.}
\end{supplement}

% imsref loaded by linak, 2015-09-28 14:15:00
%
% imsref loaded by linak, 2015-09-28 14:50:48

\printaddresses
\end{document}